# Taxonomy for Engineered Living Materials

Andrés Díaz Lantada[1,*], Jan G. Korvink[2], and Monsur Islam[2,*]

[1] *Department of Mechanical Engineering, Universidad Politécnica de Madrid, José Gutiérrez Abascal 2, 28006 Madrid, Spain*

[2] *Institute of Microstructure Technology, Karlsruhe Institute of Technology, Hermann-von-Helmholtz-Platz 1, 76344 Eggenstein-Leopoldshafen, Germany*

*Corresponding authors: andres.diaz@upm.es (A.D.L.); monsur.islam@kit.edu, monsurislam79@gmail.com (M.I.)

**Abstract**

Engineered living materials (ELMs) are the most relevant contemporary revolution in materials science and engineering. These ELMs aim to outperform current examples of "smart", active or multifunctional materials, enabling countless industrial and societal applications. The "living" materials facilitate unique properties, including autonomy, intelligent responses, self-repair, and even self-replication. Within this dawning field, most reviews and documents have divided ELMs into biological ELMs (bio-ELMs), which are solely made of cells, and hybrid living materials (HLMs), which consist of abiotic chassis and living cells. Considering that the most relevant feature of living material is that they are made of (or include) living cell colonies and microorganisms, we consider that ELMs should be classified and presented differently, more related to life taxonomies than to materials science disciplines. Towards solving the current need for the classification of ELMs, this study presents the first complete proposal of taxonomy for these ELMs. Here, life taxonomies and materials classifications are hybridized hierarchically. Once the proposed taxonomy is explained, its applicability is illustrated by classifying several examples of bio-ELMs and HLMs, and its utility for guiding research in this field is analyzed. Finally, possible modifications and improvements are discussed, and a call for collaboration is launched for progressing in this complex and multidisciplinary field.

## 1. Introduction

Taxonomies are schemes for classification, typically hierarchical, which help organize and index knowledge and research fields. Apart from classifying what already exists, taxonomies can be used to foresee and organize what may still be developed or created, especially in the cases of taxonomies dealing with novel research fields. Life sciences researchers have been trying to organize life since antiquity, although modern taxonomy starts with the works of Linnaeus [1] and with subsequent improvements aimed at incorporating evolution to it [2,3]. Even today, the task of organizing life is overwhelming and a source of intense and fruitful debate [4], also illustrated by the recent development of the PhyloCode [5] as an alternative to more traditional taxonomies.

In any case, taxonomies for life will surely need to be reformulated to adequately account for creations from the realm of synthetic biology. For more than a century (the term "synthetic biology" dates back to around 1910 [6]), life and technological scientists have joined efforts to regenerate or "engineer" damaged tissues, to edit pieces of genetic material, to artificially promote immune responses, and to create synthetic cells from basic constituents, to cite some examples, pursuing advanced therapies and a better understanding of the fundamental mechanisms of life. These research directions have set the foundations for the synthetic biology field, which also deals with the design and materialization of living machines [7] and living materials [8,9], the central topic of this study.

Engineered living materials (ELMs) are considered the most relevant contemporary revolution in materials science and engineering. These ELMs aim to outperform current examples of "smart", active or multifunctional materials and, hence, enable countless industrial and societal applications, in which the possibility of counting with materials that are "alive" improves autonomy, intelligent responses, self-repair, and even self-replication. In the last two decades, the first examples of true living materials have been achieved, normally incorporating living cells to scaffolding structures or entirely by self-assembly of living cells. Respectively, these two approaches are commonly referred to bottom-up synthesis of "biological engineered living materials" (bio-ELMs) and top-down creation of "hybrid living materials" (HLMs) (Figure 1) [10]. The more relevant pioneering endeavors and results of ELMs, both HLMs and bio-ELMs, have been recently reviewed [9,11–13]. The emergent ELMs field is already populated with hundreds of examples demanding a taxonomical organization to understand better what has been obtained and what may still be invented or created.

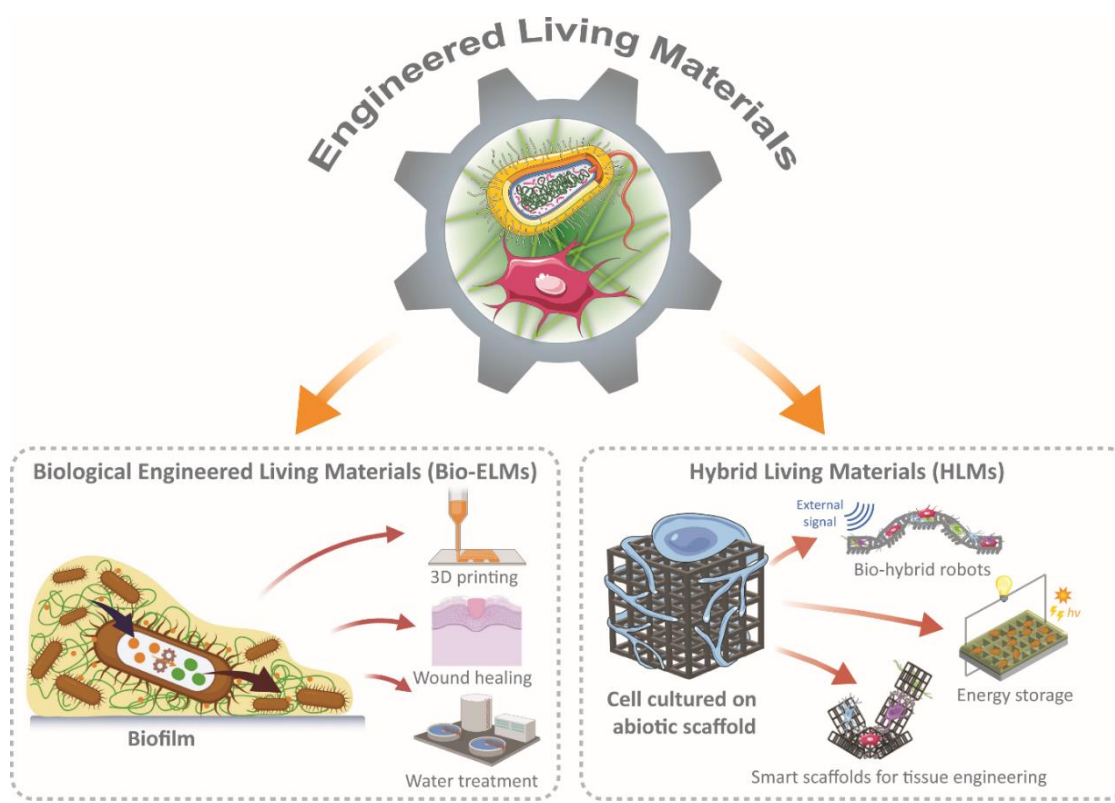

**Figure 1:** Scheme of engineered living materials and its categorization into biological engineered living materials (bio-ELMs) and hybrid living materials (HLMs). Few examples of bio-ELMs and HLMs are also depicted. It should be noted that the applications of ELMs are not limited to the ones represented here.

Among inspiring efforts for organizing this dawning field, Srubard III proposes a taxonomy for ELMs research [14], in which scale, design, organism type, material properties, and application are used as taxons. That study provides an excellent introduction to ELMs in general and highlights related research areas and trends. Nevertheless, such taxonomy is not *sensu stricto* for ELMs, but for the actual research field. For example, a specific living material may have several functions, and using "function" as taxon for ELMs would be, metaphorically, like using "profession" for classifying the species of the Homo genus. Design and application are interesting taxons for gathering researchers from different realms (i.e., space engineers, materials scientists, textile designers, to name a few) and inspiring them to join forces for transformative research in the ELMs field. But they are not essential features of engineered living materials for their univocal classification. Hierarchically speaking, in our view, organism type should be the fundamental taxon, not the third one, if a taxonomy for ELMs is developed.

Consequently, classifying a research field differs from classifying its outputs, the actual living materials, as we intend here.

Whether ELMs should be considered life or not and the related ethical implications are topics beyond the purpose of this study. However, we believe that it is plausible and even probable that, if research efforts deploy as expected, ELMs will have to be considered living entities and incorporated into the taxonomy of life. Most reviews on ELMs have organized the field considering the HLM / bio-ELM areas and the types of materials used as scaffolds or abiotic chassis for the case of HLMs. The European Union Commission also employed this approach in the recently launched EIC Pathfinder Challenge on "Engineered Living Materials" [10]. Nevertheless, considering that the most relevant feature of living materials is that they are made of or include living cell colonies and microorganisms, we consider that ELMs should be classified and presented differently, more related to life taxonomies than materials science disciplines.

To approach solving this need for classification, to the authors' best knowledge, this study presents the first complete proposal of taxonomy for ELMs, in which life taxonomies and materials classifications are hybridized hierarchically. Once the proposed taxonomy is explained, its applicability is illustrated by classifying several examples of bio-ELMs and HLMs, and its utility for guiding research in this field is analyzed. Finally, possible modifications and improvements are discussed, and a call for collaboration is launched, as needed for progressing in such a complex and multidisciplinary field.

## 2. Proposed taxonomy
### 2.1 Rationale for the taxonomy proposal

If living materials are considered life, the higher rank taxons (domains, kingdoms) should take inspiration from life's taxonomy. In fact, the most relevant and distinctive feature of any ELM is the living constituents that make it "alive". On the other hand, the lower rank taxons (classes and below) would depend on the materials' types used for the living materials' "skeletons" or chassis, abiotic in the case of HLMs, providing structural support to the living cell colonies. In this way, we propose merging life's taxonomy with common materials classifications.

Even from a functional or an application perspective, cell type is the most fundamental feature for living materials and, therefore, should constitute the higher rank. For instance, the living material employed for a building cover, capable of gathering energy from sunlight, relies on being formed by prokaryotic or eukaryotic cells, whether their support is a carbon mesh, chainmail, or an interwoven glass fiber. At the same time, the capability of living material for surviving in remote locations and under extreme pressures or temperatures, for energy or materials production purposes related to space colonization, depends on including extremophilic archaeas or prokaryotes, more than on counting with a ceramic, carbon or metallic structure. To cite another example, self-contractile microbiobots can be made of hydrogels, soft elastomers, or 3D printed mechanisms using a wide set of materials families for their abiotic chassis, but their actual autonomous motion and life expectancy depend on the use of mammal cardiomyocytes and a related life-supporting extracellular matrix, biochemical factors, and nutrients. Changing the cardiomyocytes by electrically controlled musculoskeletal cells, or incorporating neural cells for future cognitive tasks, may affect the performance and potentials of such microbiobots much more than modifying the materials used as chassis.

On the basis of the above, the following subsections present the taxonomy, from the higher rank to the lower rank taxons, before analyzing its viability and utility for classification and discovery purposes.

## 2.2 Domains, super-kingdoms (or empires), and kingdoms inspired by life's taxonomies

Our proposal for the higher rank taxons takes inspiration from the more widespread taxonomy for life, which uses three domains and six kingdoms (Woese's system [15]). Accordingly, we propose archaeal, bacterial, and eukaryotic living materials as basic domains and complement them with an additional domain for those with synthetic cells as constituents. Besides, chimeric creations that fail in nature may find a successful application and prove viable in synthetic biology, for which we add a cross-domain for living materials created by combining cells from other domains. The same principles apply to the kingdoms (archaebacterial, eubacterial, protist, fungal, vegetal, animal) ELMs, to which we also add a kingdom for ELMs with artificial cells and a cross-kingdom.

Between the domains and kingdoms, we consider the possibility of two superkingdoms: biological engineered living materials (made only of cells) and hybrid living materials (combining cells with an abiotic structure). As previously mentioned, these are already seen and put forward in different documents as the two main types of ELMs. Including this division in the taxonomy may be useful for integrating an already standard classification of ELMs and because scaffold-free living materials (bio-ELMs) may still evolve in unexpected directions, while HLMs seem already more established. A similar revolution is now taking place in the related field of tissue engineering, in which scaffold-free approaches are gaining research attention after decades of scaffold-based developments, and synergies among both categories are being increasingly explored [16]. Nonetheless, the long-term utility of such superkingdoms and a possible simplification of the taxonomy are discussed in Section 4. The taxonomy for the higher-rank taxons is presented below in Figure 2.

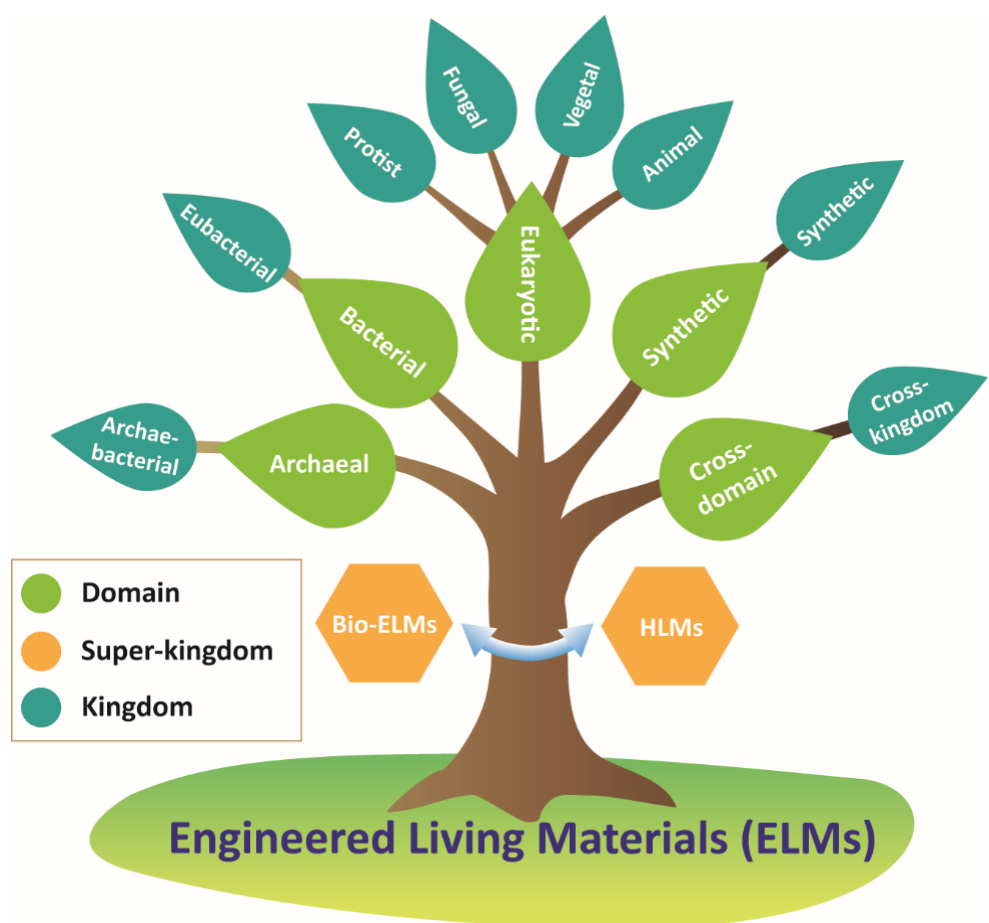

**Figure 2:** The taxonomy of ELMs based on the higher-rank taxons, showing the domains, super-kingdoms, and kingdoms of ELMs.

## 2.3 Phyla, classes, families, and species inspired by materials classifications

Proposed phyla act as the bridge between the higher ranks, linked to the living entities, and the lower ranks, connected to the materials used as extracellular matrices or chassis for the ELMs. Considering that life is divided into phyla depending on the spatial configuration of their cells and that all kinds of materials can also be grouped according to their topology, we propose phyla based on dimensional features for the living materials taxonomy, as illustrated in Figure 3.

Consequently, n-dimensional phyla lead to 0D, 1D, 2D, and 3D living materials. If the materials, apart from being alive, can move and interact with the environment through autonomous or controlled shape-morphing or through purposely designed degradation, the living material can be considered 4D ELMs. Those living materials with multi-scale self-replicating morphology, recursive definition, or non-integer dimensions are fractal ELMs. Cross-phylum living materials and derived systems may combine, for example, two-dimensional biofilms with three-dimensional scaffolds.

In our opinion, these dimension-derived phyla will be useful, as both living tissues and traditional non-living materials are classifiable through their dimensional features. Besides, the studies on biological materials, in many cases, describe and model tissues as 1D fibers working under traction, as 2D shells and meshes, or as 3D constructs. 2D / 3D classifications have been also proved useful for organizing another emergent field, metamaterials, into 3D metamaterials and quasi-2D metasurfaces. 3D and 4D terms are also applied to classifying static versus dynamic printed materials, which are also opening new horizons in materials science.

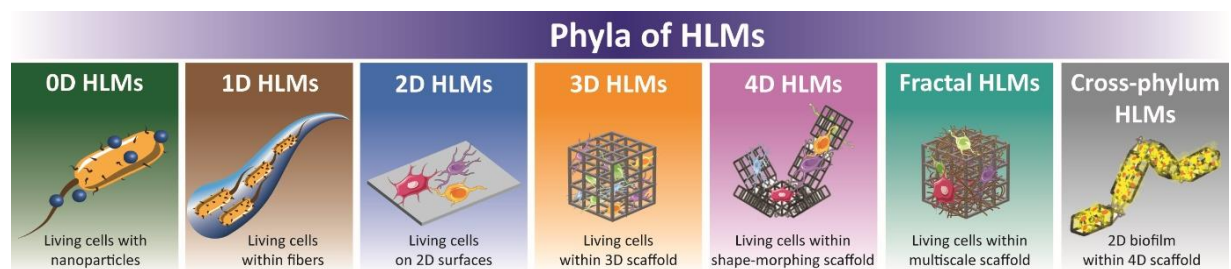

**Figure 3:** Taxonomy of ELMs, particularly HLMs, based on their phylum.

Regarding the proposed classes for the taxonomy, the actual type of material used as extracellular matrices or abiotic chassis is considered. This applies mainly to HLMs in our first taxonomy proposal but could be considered for all kinds of ELMs, if some minor modifications are included, as discussed in Section 4. The common materials classification depending on the chemical bonds and composition, which determine the mechanical, electrical, and thermal properties, is used here for the classes of living materials. This leads to metallic, ceramic, polymeric, carbon, and composite ELMs (here HLMs), which in short can be referred to as living metals, living ceramics, living polymers, living carbons, and living composites, respectively.

Families are organized by further dividing the classes of materials, used as extracellular matrices or chassis, by employing the common subdivisions used in materials science. For example, Living metals are divided into families, including living steels, living Cu-alloys, living Ni-alloys, to name a few. The same applies to the families of other living materials classes, presented in detail in Figure 4. As for the species, an additional level of detail is expected. For instance, the living CNTs family can be further categorized into living SWCNTs and MWCNTs, and living thermoplastic polymer can include specific species of living PLA, living PCL, or living PLGA. The use of specific cell colonies from concrete living species may also lead to different species within the same family.

Evidently, the species are too many for being enumerated here, and additional efforts are needed to completely detail the whole taxonomy and related nomenclature, as further discussed in the final sections. Before entering discussions and future proposals, the next section demonstrates the applicability and puts forward the utility of the proposed living materials taxonomy.

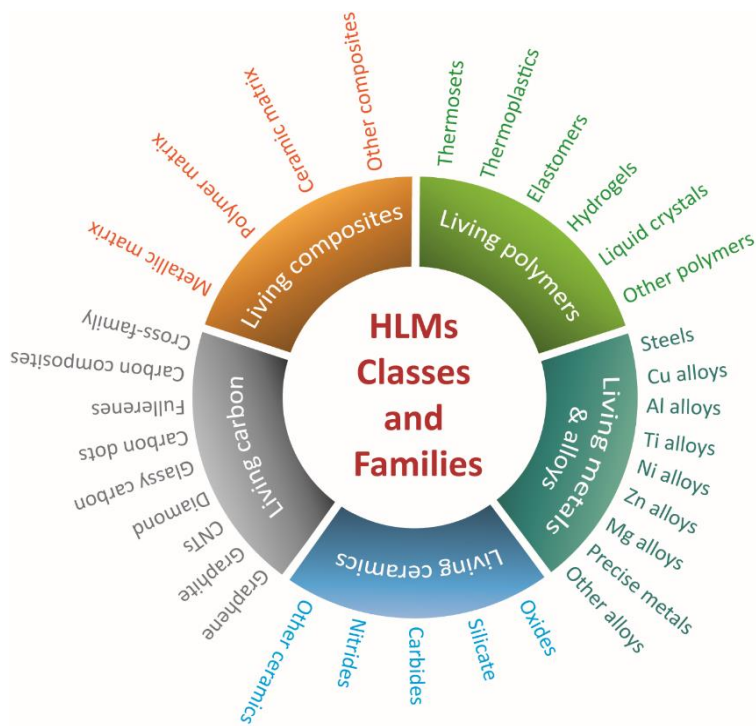

**Figure 4.** Taxonomy for ELMs: families for different classes of HLMs.

## 3. Applicability and utility of the taxonomy

### 3.1 Classification of pioneering examples of living materials

A total of 75 examples of ELMs have been categorized according to our taxonomy proposal to validate their applicability and utility, as summarized in Table 1. The following considerations are taken into account, for constructing Table 1 according to our proposed taxonomy, together with our previously described rationale and taxonomic structure:

Within the superkingdom of bio-ELMs [17,18], we include the ELMs, which are made of only cells or cells and cells-generated extracellular matrices, without the employment of an external scaffold or abiotic chassis. When ELMs are constructed using scaffolding structures by seeding cells within them or through direct interaction and self-assembly of cells and biomaterials, we classify them as hybrid living materials or HLMs. In some cases, cells cultured in suspension (usually for materials production purposes) are included as biological ELMs, while cells cultured upon biofilms of biomaterials are HLMs.

We have not included tissue engineering scaffolds as ELMs for different reasons. Although tissue engineering scaffolds are historical precursors and relatives to ELMs, they are typically employed as structural elements, and the role of cells within them is to improve their biological performance. With literally thousands of examples of tissue engineering scaffolds populating the fields of tissue engineering and biofabrication, they constitute a research area on their own. Besides, in the vast majority of cases, they lack most of the representative features of true ELMs, including intelligent responses, autonomous motion or interactions with the environment, self-sensing abilities, self-repair capabilities, and self-replication. In some exceptional cases, we have included tissue engineering scaffolds in Table 1 due to their more apparent connection with ELMs or their pioneering examples of biohybrid combinations (i.e. new type of chassis or innovative combinations of biological entities across different domains).

In addition, the term "artificial cell" or "synthetic cell" is emergent and can represent different entities, like integral biological imitators, cell-like structures with biological functions, engineered existing cells with modified functions, or inert bio-templated cell-like structures, to cite a few [19,20]. In some cases, these so-called "artificial cells" are inert polymeric bubbles or chambers, which may be used for drug delivery or for distracting pathogens but cannot be considered true living entities or living materials because they lack the basic functions of life. Engineered cells, using common synthetic biology tools like m-RNA or CRISP-PR, are sometimes called artificial cells. However, a human being receiving an m-RNA-based vaccine or born from an engineered embryo to avoid a congenital disease would still be a human being. Applying a similar rationing scheme, an ELM based on edited animal cells would still be an "animal ELM", and an ELM benefiting from modified bacteria should still be considered an "eubacterial ELM". In consequence, our list of ELMs made of artificial cells does not include biotemplated inert materials or engineered living cells employing state-of-the-art genetic engineering technologies. We opt for completely synthetic cells, in which artificially produced microstructural entities are loaded with synthetic genes or organelles for autonomously interacting with the environment and count with basic living features, like self-replication, self-sensing abilities, independent motion, and collective interactions. Although this decision limits the number of existing examples of ELMs made of true artificial cells, we believe that it agrees with recent definitions from relevant research calls [10] and studies [21–23] and that several examples of ELMs with artificial cells will be developed in the short-term.

Considering the above, pioneering examples of ELMs classified according to the taxonomy are presented in the following pages (Table 1). Its utility is further analyzed in section 3.2.

| Pioneering examples of engineered living materials (ELMs) | | Classification according to proposed taxonomy | | | | | |
|---|---|---|---|---|---|---|---|
| Described engineered living materials (ELMs) | Purpose / Application field | Domain | Superkingdom | Kingdom | Phylum | Class | Familiy |
| Archaeal cultures for polymeric production [24] | Materials production | Archaeal | Biological ELMs | Archaebacterial ELMs | 3D ELMs | Living polymers | Living thermoplastics |
| Archaeal cultures for gold production [25] | Materials production | Archaeal | Biological ELMs | Archaebacterial ELMs | 3D ELMs | Living metals | Living precious metals |
| Archaeal cable-like structures upon biofilms [26] | Smart surfaces and structures | Archaeal | Hybrid living materials | Archaebacterial ELMs | 1D/2D ELMs | Living polymers | Living elastomers |
| Cellulose produced by bacteria in culture [27] | Materials production | Bacterial | Biological ELMs | Eubacterial ELMs | 3D ELMs | Living polymers | Living thermoplastics |
| Cellulose produced by bacteria in culture [28] | Materials production | Bacterial | Biological ELMs | Eubacterial ELMs | 3D ELMs | Living polymers | Living thermoplastics |
| 3D printed self-healing bacterial biofilms [29] | Materials production | Bacterial | Biological ELMs | Eubacterial ELMs | 3D ELMs | Living polymers | Living elastomers |
| Genetically programmable self-regen. bacterial hydrogels [30] | Tissue engineering | Bacterial | Biological ELMs | Eubacterial ELMs | 3D ELMs | Living polymers | Living hydrogels |
| Bacterial-biomineralized inorganic substrates [31] | Materials production | Bacterial | Biological ELMs | Eubacterial ELMs | 3D ELMs | Living ceramics | Living concrete |
| Extracellular matrix created by bacteria in culture [32] | Materials production | Bacterial | Biological ELMs | Eubacterial ELMs | 2D ELMs | Living polymers | Living hydrogels |
| Bacterial cultures for polymeric production [33] | Materials production | Bacterial | Biological ELMs | Eubacterial ELMs | 2D ELMs | Living polymers | Living thermoplastics |
| Bacterial cultures for polymeric production [34] | Materials production | Bacterial | Biological ELMs | Eubacterial ELMs | 2D ELMs | Living polymers | Living thermoplastics |
| Bacterial biofilms for gold production [25] | Materials production | Bacterial | Biological ELMs | Eubacterial ELMs | 3D ELMs | Living metals | Living precious metals |
| Bacterial cultures for gold nanoparticles production [35] | Materials production | Bacterial | Biological ELMs | Eubacterial ELMs | 0D ELMs | Living metals | Living precious metals |
| Bacterial cultures for silver nanoparticles production [36] | Materials production | Bacterial | Biological ELMs | Eubacterial ELMs | 0D ELMs | Living metals | Living precious metals |
| Bacterial biofilms with nanoparticles and quantum dots [37] | Materials production | Bacterial | Hybrid living materials | Eubacterial ELMs | Fractal | Living metals | Living precious metals |
| Sand-hydrogel scaffolds with photosynthetic cyanobacteria [38] | Smart buildings | Bacterial | Hybrid living materials | Eubacterial ELMs | 3D ELMs | Living composites | Living polym. matrix comp. |
| Bacteria cultured on polymeric scaffolds growing ceramics [39] | Materials production | Bacterial | Hybrid living materials | Eubacterial ELMs | 3D ELMs | Living composites | Living ceram. matrix comp. |
| Self-healing bacterial loaded concrete [40] | Smart buildings | Bacterial | Hybrid living materials | Eubacterial ELMs | 3D ELMs | Living ceramics | Living concrete |
| Hydrogel hosting genetically programmed bacteria [41] | Smart surfaces and structures | Bacterial | Hybrid living materials | Eubacterial ELMs | 3D ELMs | Living polymers | Living hydrogels |
| 3D bioprinted hydrogels with programmed bacteria [42] | Smart surfaces and structures | Bacterial | Hybrid living materials | Eubacterial ELMs | 3D ELMs | Living polymers | Living hydrogels |
| 2D and 3D bioprinted hydrogels encapsulating bacteria [43] | Smart surfaces and structures | Bacterial | Hybrid living materials | Eubacterial ELMs | 3D ELMs | Living polymers | Living hydrogels |
| Smart grippers with sensing bacteria [44] | Robotics | Bacterial | Hybrid living materials | Eubacterial ELMs | 3D ELMs | Living polymers | Living elastomers |
| Multi-layered polymeric biofilm with embedded bacteria [45] | Smart surfaces and structures | Bacterial | Hybrid living materials | Eubacterial ELMs | 3D ELMs | Living polymers | Living elastomers |
| Bacteria grown on marine agar synthesize silk proteins [46] | Materials production | Bacterial | Hybrid living materials | Eubacterial ELMs | 3D ELMs | Living polymers | Living elastomers |
| Penicilin-producing living surfaces [47] | Biotechnology and bioprocessing | Bacterial | Hybrid living materials | Eubacterial ELMs | 2D ELMs | Living polymers | Living hydrogels |
| 3D bioprinted photosynthetic cianobacteria [48] | Energy production | Bacterial | Hybrid living materials | Eubacterial ELMs | 2D ELMs | Living polymers | Living hydrogels |
| Polymeric films loaded with water-responsive bacterial spores [49] | Smart surfaces and structures | Bacterial | Hybrid living materials | Eubacterial ELMs | 2D ELMs | Living polymers | Living elastomers |
| Biosensing alginate beads with bacterial colonies [50] | Smart surfaces and structures | Bacterial | Hybrid living materials | Eubacterial ELMs | 2D ELMs | Living polymers | Living elastomers |
| Nanofibrous webs hosting bacteria [51] | Biotechnology and bioprocessing | Bacterial | Hybrid living materials | Eubacterial ELMs | 2D ELMs | Living polymers | Living elastomers |
| Bacteria patterned onto textiles and polymers [52] | Smart surfaces and structures | Bacterial | Hybrid living materials | Eubacterial ELMs | 2D ELMs | Living polymers | Living thermoplastics |
| Self-assembled graphene-bacterial biofilms [53] | Smart surfaces and structures | Bacterial | Hybrid living materials | Eubacterial ELMs | 2D ELMs | Living carbons | Living graphene |
| Graphene biofilms with electroactive bacteria [54] | Energy production | Bacterial | Hybrid living materials | Eubacterial ELMs | 2D ELMs | Living carbons | Living graphene |
| Graphene-CNT bioflims with built-in bacteria [55] | Energy production | Bacterial | Hybrid living materials | Eubacterial ELMs | 2D ELMs | Living carbons | Other living carbons |
| Polymeric fibers and meshes loaded with bacteria [56] | Biotechnology and bioprocessing | Bacterial | Hybrid living materials | Eubacterial ELMs | 1D ELMs | Living polymers | Living thermoplastics |
| Biogenic gold nanoparticles with radiation-resistant bacteria [57] | Biotechnology and bioprocessing | Bacterial | Hybrid living materials | Eubacterial ELMs | 0D ELMs | Living metals | Living precious metals |
| C-dots synthesized by bacteria upon GO films [58] | Biotechnology and bioprocessing | Bacterial | Hybrid living materials | Eubacterial ELMs | 0D ELMs | Living carbons | Other living carbons |
| Multi-scale carbon structures encapsulating cells [59] | Tissue engineering | Eukaryotic | Hybrid living materials | Animal ELMs | Fractal | Living carbons | Living glassy carbon |
| Microencapsulated mammalian cells in natural polymers [60] | Tissue engineering | Eukaryotic | Hybrid living materials | Animal ELMs | 3D ELMs | Living polymers | Living elastomers |
| Self-contractile PDMS with cardiomyocytes [61] | Robotics | Eukaryotic | Hybrid living materials | Animal ELMs | 3D ELMs | Living polymers | Living elastomers |
| Collagen structure with musculoskeletal tissue [62] | Robotics | Eukaryotic | Hybrid living materials | Animal ELMs | 3D ELMs | Living polymers | Living elastomers |
| Biological machines with musculoskeletal cells [63] | Robotics | Eukaryotic | Hybrid living materials | Animal ELMs | 3D ELMs | Living polymers | Living hydrogels |
| Bioencapsulated animal cells in silica matrices [64] | Smart surfaces and structures | Eukaryotic | Hybrid living materials | Animal ELMs | 3D ELMs | Living ceramics | Living silica |
| Cardiomyocytes on elastomeric body with gold skeleton [65] | Robotics | Eukaryotic | Hybrid living materials | Animal ELMs | 3D ELMs | Living composites | Living polym. matrix comp. |
| PDMS thin films with cardiomyocytes [66] | Robotics | Eukaryotic | Hybrid living materials | Animal ELMs | 2D ELMs | Living polymers | Living elastomers |
| PDMS medusoids with living tissues [67] | Robotics | Eukaryotic | Hybrid living materials | Animal ELMs | 2D ELMs | Living polymers | Living elastomers |
| Muskuloskeletal cells attached to hair [68] | Robotics | Eukaryotic | Hybrid living materials | Animal ELMs | 1D ELMs | Living polymers | Living elastomers |
| Mycelium based composites (funghi into organic substrate) [69] | Materials production | Eukaryotic | Hybrid living materials | Fungal ELMs | 3D ELMs | Living polymers | Other living pols. (fibers) |
| Bioprinted bioinks with baker's yeast [70] | Biotechnology and bioprocessing | Eukaryotic | Hybrid living materials | Fungal ELMs | 3D ELMs | Living polymers | Living thermoplastics |
| Mycelium based composites (funghi into inorganic structure) [71] | Smart buildings | Eukaryotic | Hybrid living materials | Fungal ELMs | 3D ELMs | Living polymers | Living thermoplastics |
| Swelling-responsive hydrogels with yeast [72] | Smart surfaces and structures | Eukaryotic | Hybrid living materials | Fungal ELMs | 3D ELMs | Living polymers | Living hydrogels |
| Yeast-CNT bionic nanocomposite [73] | Smart surfaces and structures | Eukaryotic | Hybrid living materials | Fungal ELMs | 3D ELMs | Living carbons | Living CNTs |
| Yeast-graphene bionic nanocomposite [74] | Smart surfaces and structures | Eukaryotic | Hybrid living materials | Fungal ELMs | 3D ELMs | Living carbons | Living graphene |
| Silica-alginate-funghi biocomposites [75] | Biotechnology and bioprocessing | Eukaryotic | Hybrid living materials | Fungal ELMs | 3D ELMs | Living composites | Living polym. matrix comp. |
| Polymeric layers with *Penicillium roqueforti* [76] | Smart surfaces and structures | Eukaryotic | Hybrid living materials | Fungal ELMs | 2D ELMs | Living polymers | Living elastomers |
| Algae-laden hydrogel scaffolds [77] | Tissue engineering | Eukaryotic | Hybrid living materials | Protist ELMs | 3D ELMs | Living polymers | Living hydrogels |
| 3D printed corals hosting microalgae [78] | Energy production | Eukaryotic | Hybrid living materials | Protist ELMs | 3D ELMs | Living composites | Living ceram. matrix comp. |
| Bioprinted photosynthetic living materials [79] | Energy production | Eukaryotic | Hybrid living materials | Protist ELMs | 2D ELMs | Living composites | Living polym. matrix comp. |
| Trees interwoven with scaffolding structures [80] | Smart buildings | Eukaryotic | Hybrid living materials | Vegetal ELMs | Fractal | Living metals | Living steel |
| Plant cell-laden hydrogel scaffolds [81] | Biotechnology and bioprocessing | Eukaryotic | Hybrid living materials | Vegetal ELMs | 3D ELMs | Living polymers | Living hydrogels |
| Plant cells in silica matrices [82] | Smart surfaces and structures | Eukaryotic | Hybrid living materials | Vegetal ELMs | 3D ELMs | Living ceramics | Living silica |
| Bioencapsulated plant cells in silica matrices [64] | Smart surfaces and structures | Eukaryotic | Hybrid living materials | Vegetal ELMs | 3D ELMs | Living ceramics | Living silica |
| Self-assembled DNA nanotubes and artificial cells [83] | Synthetic biology | Synthetic cells | Biological ELMs | Artificial cells ELMs | 3D ELMs | Living polymers | Other living polymers |
| Photosynthetic membranes with synthetic chloroplasts [84] | Biotechnology and bioprocessing | Synthetic cells | Biological ELMs | Artificial cells ELMs | 2D ELMs | Living polymers | Other living polymers |
| Protocellular models producing sugar and stimulating bacteria [85] | Biotechnology and bioprocessing | Synthetic cells | Biological ELMs | Artificial cells ELMs | 0D ELMs | Living polymers | Other living polymers |
| Protocells for the synthesis of ATP [86] | Biotechnology and bioprocessing | Synthetic cells | Biological ELMs | Artificial cells ELMs | 0D ELMs | Living polymers | Other living polymers |
| Mushroom with cyanobacteria and graphene nanoribbons [87] | Energy production | Cross-domain | Hybrid living materials | Cross-kingdom ELMs | Fractal | Living carbons | Living graphene |
| Lichen-inspired latex with microalgae and cyanobacteria [88] | Biotechnology and bioprocessing | Cross-domain | Hybrid living materials | Cross-kingdom ELMs | 3D ELMs | Living polymers | Living elastomers |
| Algae, yeast and bacteria encapsulated in bioprintable hydrogel [89] | Smart surfaces and structures | Cross-domain | Hybrid living materials | Cross-kingdom ELMs | 3D ELMs | Living polymers | Living elastomers |
| Synergic smart materials production by bacteria and yeast [90] | Materials production | Cross-domain | Hybrid living materials | Cross-kingdom ELMs | 3D ELMs | Living polymers | Living thermoplastics |
| Bacterial nanocellulose scaffolds for endothelial cultures [91] | Tissue engineering | Cross-domain | Hybrid living materials | Cross-kingdom ELMs | 3D ELMs | Living polymers | Living thermoplastics |
| Cell scaffolds with antibiotic-secreting bacteria [92] | Tissue engineering | Cross-domain | Hybrid living materials | Cross-kingdom ELMs | 3D ELMs | Living polymers | Living hydrogels |
| Bacteria-laden gels for stem cell engineering [93] | Tissue engineering | Cross-domain | Hybrid living materials | Cross-kingdom ELMs | 3D ELMs | Living polymers | Living hydrogels |
| Biomaterials with interactions among hMSCs and bacteria [94] | Tissue engineering | Cross-domain | Hybrid living materials | Cross-kingdom ELMs | 2D ELMs | Living polymers | Living hydrogels |
| Hydrogel surfaces with bacteria to trigger cell adhesion [95] | Tissue engineering | Cross-domain | Hybrid living materials | Cross-kingdom ELMs | 2D ELMs | Living polymers | Living hydrogels |
| Soft matter created by cooperation of microalgae and bacteria [96] | Materials production | Cross-domain | Hybrid living materials | Cross-kingdom ELMs | 2D ELMs | Living polymers | Living thermoplastics |

**Table 1.** Application of the taxonomy to a collection of engineered living materials described in recent literature.

## 3.2 Finding gaps in the living materials portfolio and guiding research efforts

Apart from helping to classify what already exists, taxonomies can be applied to analyze what could exist, especially in a field connected to synthetic biology. This will further help propose new research directions. After classifying the selected pioneering examples of ELMs according to the proposed taxonomy, the most populated domains, kingdoms, and classes are analyzed and summarized in Figure 5 and 6.

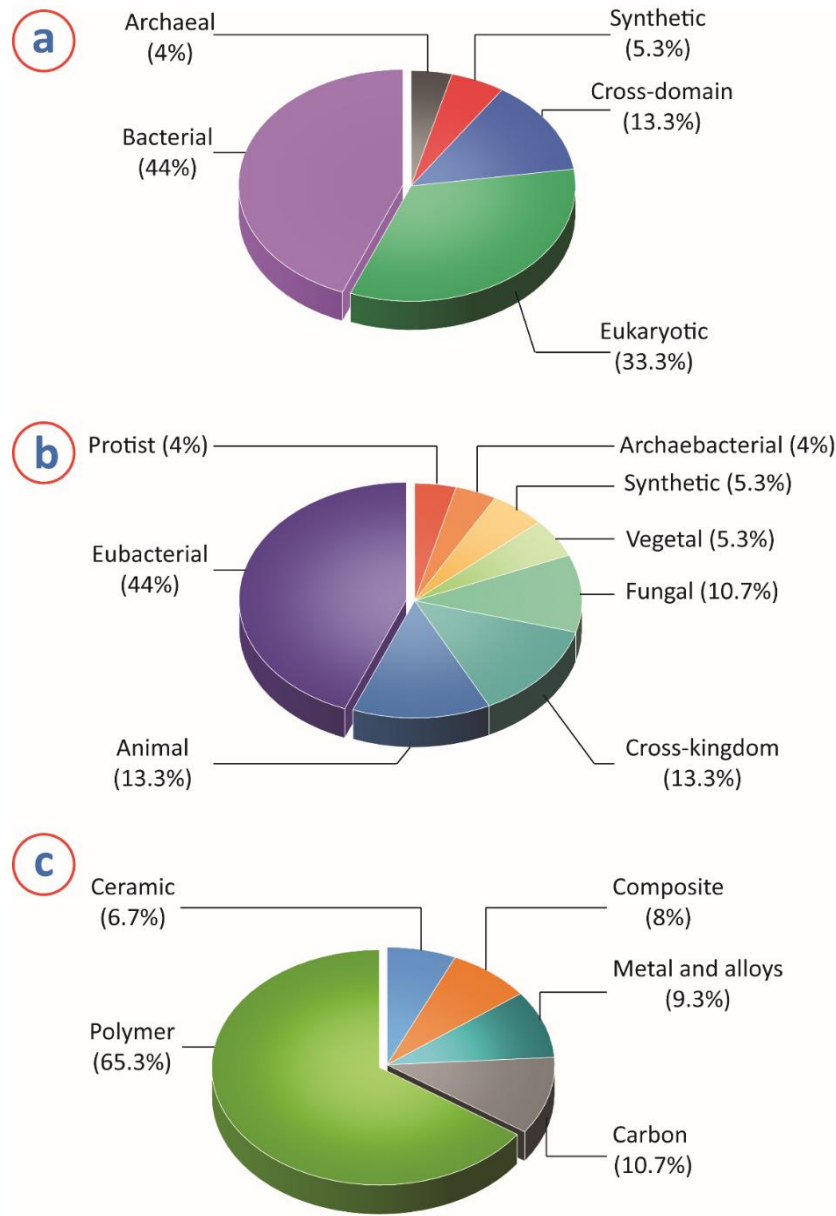

**Figure 5.** Engineered living materials by: a) domain, b) kingdom, and c) class, ordered by number of examples found in the literature search according to Table 1.

According to the gathered examples and data, bacterial ELMs account for 44% of examples of living materials, while eukaryotic ELMs correspond to 33% and archaeal ELMs to 4%. The industrial relevance of bacterial cultures for white, green, and blue biotechnologies is probably responsible for this prevalence, while several examples of eukaryotic ELMs are connected to fields like tissue engineering and biohybrid robotics. Up to now, the potential of archaea has not been fully exploited, probably due to their less studied nature. Their extremophilic properties may find relevant applications linked to space exploration and developing resources and raw materials in remote and extremely harsh environments.

ELMs with completely synthetic artificial cells, with 5% of examples, are still underrepresented, considering their extraordinary potentials. We assume that this situation will significantly change soon. Interestingly a remarkable 13% of studies deal with cross-domain ELMs, in which bacterial and eukaryotic cells co-exist and perform symbiotic functions.

Regarding the employment of an abiotic chassis or not, HLMs constitute 77% of the analyzed examples, while biological ELMs without synthetic scaffolding structures account for the remaining 23%. Within HLMs, most chassis (65.3%) are polymeric, followed by carbon-based materials (10.7%), metals and alloys (9.3%), composites (8%), and ceramics (6.7%). The distribution of the different ELM kingdoms and classes is presented in Figure 6. Considering polymers, the interesting properties of hydrogels for cell culture applications make them a usual choice (30% within polymers), despite their very poor mechanical properties. Elastomeric materials like PDMS are also employed similarly to that of hydrogels, although their improved properties do not allow for truly high-performance applications.

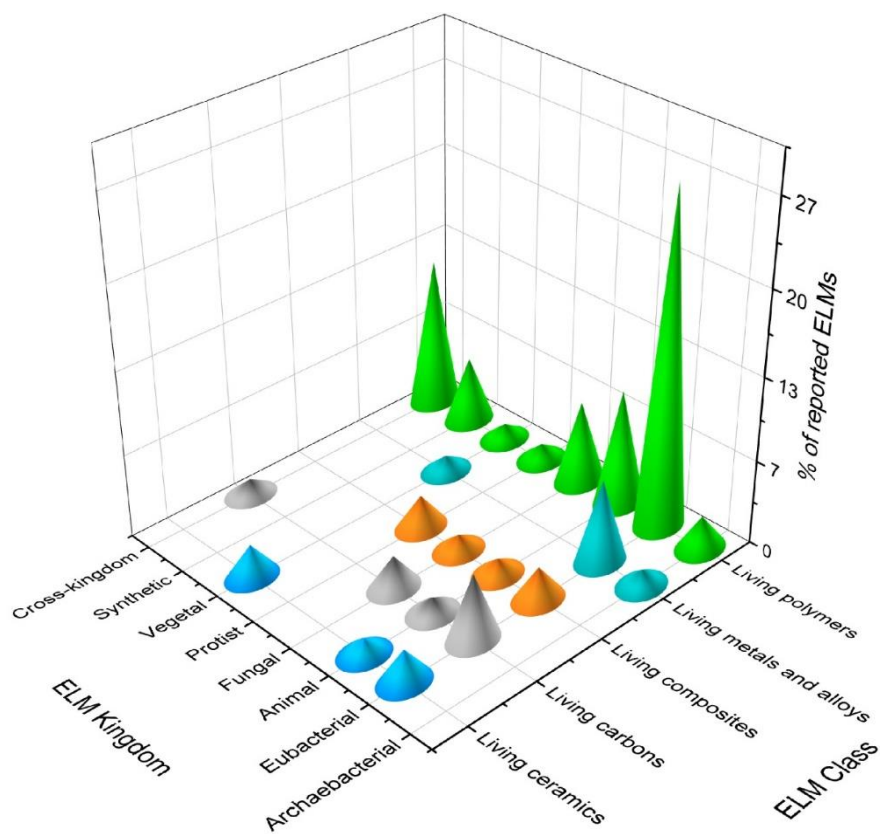

**Figure 6.** Engineered living materials by kingdoms, according to the living components, and classes, according to the materials of the abiotic chassis or extracellular matrices.

The different application fields for the developed ELMs are also considered for their industrial and social interest, even if the application itself has not been employed as a taxon. The function of the ELMs that determine the application should be considered as a texon. However, the current application fields are quite diverse. Furthermore, as it is an emerging field, numerous applications are expected to emerge in the future. Therefore, we believe it is too early for a classification considering fuction or application as a taxon. However, this is subject to further debate and discussion. Nevertheless, we attempted to categorize different broader application areas based on Table 1. The distribution is presented in Figure 6. Application areas of materials production and smart surfaces and structures share the top position with 23% of all the applications. Materials production is due to the high interest in bio-ELMs using living cells, particularly bacteria, are  for the synthesis of extracellular materials, such as cellulose nanofibers or gold/silver nanoparticles. In comparison, HLMs account for smart surfaces and structures due to the synergic activites of living cells and the abiotic chassis. However, there are growing interests in other fields, including biotechnology and bioprocessing, tissue engineering. One of the reasons is the processability of the living entities using additive manufacturing technologies, either by directly 3D printing or through assembly with a 3D printed scaffold. The fields of bio-robotics, and energy productions are also emerging slowly utilizing unique functions of specific cell lines.

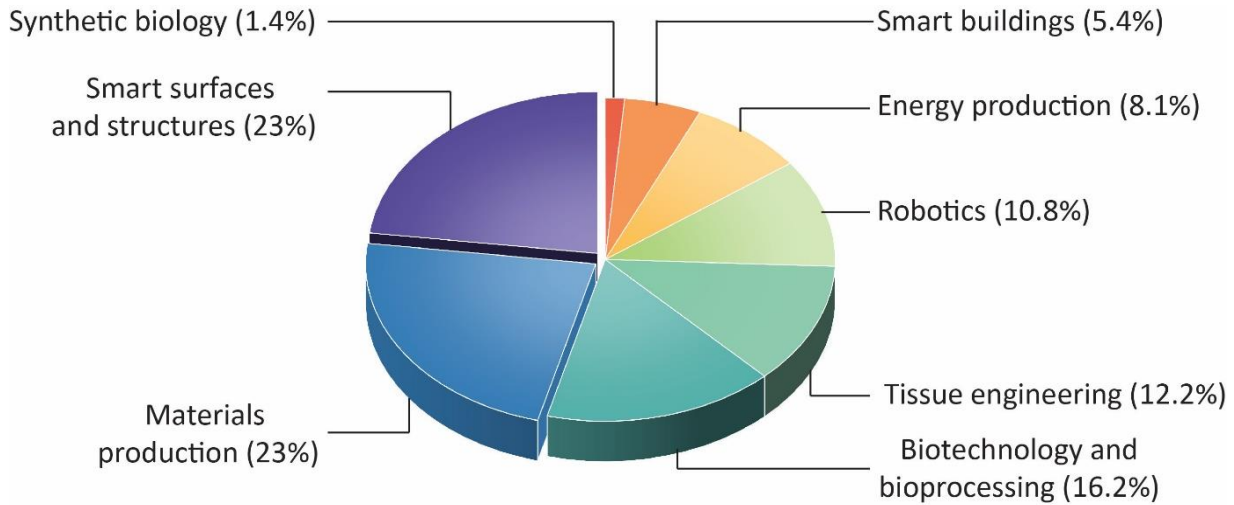

**Figure 6.** Engineered living materials by application fields.

## 4. Discussion about possible modifications and alternatives to the taxonomy

The proposed taxonomy is useful for hierarchically classifying the emergent field of ELMs in a univocal manner, and for finding unexplored combinations of living kingdoms and materials classes, to orient research efforts in the field. However, some questions for debate arise, which will need consensual responses. Further developments in ELMs, especially regarding bio-ELMs, may lead to modifications and additions to this initial taxonomy proposal. Some relevant issues that we have already detected are discussed here, with the aim of initiating fruitful debate with colleagues.

*On the utility of the bio-ELMs and HLMs super-kingdoms*

Arguably, the common division of ELMs into bio-ELMs and HLMs, with their respective bottom-up and top-down approaches, derives from the different initial methods and techniques used by the pioneering communities of biologists (usually applying bottom-up tools) and engineers (typically resorting to top-down resources). In a way, although widespread, this division is more procedimental and historical than linked to the essential features of ELMs. Therefore, it could be progressively abandoned, and the superkingdoms could be omitted in the taxonomy.

In fact, bio-ELMs, if successfully developed to achieve long-lasting living constructs, will generate their own extracellular matrices (as in normal living organisms). These extracellular matrices can perform exactly the same function as the abiotic chassis in HLMs. Following this rationale, the superkingdoms could be further modified to introduce new superkingdoms. For examples, another superkingdom "living biomaterials" could be introduced taking into account of the cell produced biological chassis (*e.g.*, proteins, polysaccharides, mineralized tissues, to name a few). New classes and families could be further intricuded as well, based on the new superkingdom.

Alternatively, future research direction could see a unification of these two superkingdoms, where the differences between bio-ELMs and HLMs could be considered reconcilable by the research community. In that case, the superkingdoms could be eliminated and a new taxonmy would be needed.

*On the completeness of the taxonomy*

On purpose, we have not yet used the intermediate and lower rank taxons of "order" and "genus" for this first approach to the taxonomy, as it will need to be completed. We understand that the ELMs field is nascent and that unforeseen methods, techniques, combinations, and synergies between engineering and biology will lead to a plethora of multi-cellular synthetic creations, for which more complex classification schemes with more taxons will be needed. If the effect of using different cell types from the same kingdom leads to clearly different ELMs, the addition of orders within the proposed families may prove useful. Even some taxons may see their ranks modified and thus enable the incorporation of new taxons changing the hierarchy. Genus may also be applied to complement the families, especially for the more populated materials families like polymers and composites. Anyhow, we believe that the global scheme is valid and robust and it can be adapted and updated as the research in this field progresses and new ELMs appear.

*On virus and prions in living materials*

Some life taxonomies have included non-cellular domains, like prionobiota (for prions) and virusobiota (for viruses) and related kingdoms. However, it is generally accepted that viruses and prions do not constitute living entities. Accordingly, they have not been proposed as domains or kingdoms for the taxonomy of living materials.

Notwithstanding this consensus, it is also true that the incorporation of prions and viruses to living materials may be transformative and lead to unexpected new entities, functions, and applications. For example, in the field of biomaterials and tissue engineering, lentiviral vectors have been used in cultured cells, with which artificial muscles have been bioengineered, for enhancing angiogenesis and bioactivity [97]. In parallel, implants coated with viral vectors encoding for osteogenic genes are being studied for enhanced bone formation in complex large-bone and cranial repairs [98], to cite some examples.

It is foreseeable that viruses will find applications within ELMs, co-cultured or incorporated to them together with other cell types. Despite this possibility, we believe that the taxonomy is prepared for these and similar circumstances without the need for adding new domains or kingdoms. Even if including viruses or prions, the living materials would be classified according to their cellular constituents. Besides, the cross-domain and cross-kingdoms options can help integrate the more complex living materials, as would be the case of multicellular chimeras functionalized with viruses or prions.

Arguably this may change in the future due to the progressive blend of frontiers between the living and non-living, to which the field of ELMs also adds new unknowns and debates. May it be the case that prions and viruses become considered living entities, the proposed taxonomy could also be updated by adding additional domains and kingdoms.

## 5. Concluding remarks

To summarize, we attempted to classify the emerging field of ELMs, taking inspirations from life taxonomies and materials classifications, particularly for HLMs. The field is still in its "infant" stage, and will require a workforce to mature in the next decades to address several global challenges of the 21st century and beyond. The proposed taxonomy will be useful for categorically employ the muscle and brain power in this field. The proposed classification will still need reformulations, as the filed gorws over time. However, With this taxonomy proposal we hope to provide a good start for a hierarchical classification of ELMs, and to promote fruitful debate leading to a more consensual, complete, univocal and long-lasting taxonomy for living materials.

**Acknowledgments**

J.G.K. and M.I. acknowledge support from the Deutsche Forschungsgemeinschaft (DFG, German Research Foundation) under Germany's Excellence Strategy via the Excellence Cluster 3D Matter Made to Order (EXC-2082/1-390761711). All the authors thank the Karlsruhe Institute of Technology and Universidad Politécnica de Madrid for their support in facilitating a safe and healthy work environment during the adverse period of the COVID-19 pandemic.